# Power Grid Transient Analysis via Open-Source Circuit Simulator: A Case Study of HVDC

Yongli Zhu
Texas A&M University
College Station, TX, USA
yzhu16@vols.utk.edu

Xiang Zhang
Purdue University
West Lafayette, IN, USA
xiang.zhang@alumni.purdue.edu

Renchang Dai
Puget Sound Energy
Bellevue, WA, USA
renchang.dai@pse.com

*Abstract*—This paper proposes an electronic circuit simulator-based method to accelerate the power system transient simulation, where the modeling of a generic HVDC (High Voltage Direct Current) system is focused. The electronic circuit simulation equations and the backward differentiation formula for numerical solving are described. Then, the circuit modeling process for power system components such as slack bus, constant power load, and HVDC are respectively illustrated. Finally, a case study is conducted on a four-bus power system to demonstrate the effectiveness of the proposed modeling and simulation method.

## I. INTRODUCTION

To date, the power utility grid has been encountering various new challenges due to the increasing penetration of nonlinear-behavior components with large power capacity and complicated structures and controllers, e.g., power inverter-based renewables (IBR), SVC (Static Var Compensator), and HVDC (high voltage direct current) systems [1]. HVDC has certain advantages over conventional AC transmission lines. For example, lower costs for rights-of-way and overhead lines; connection of asynchronous grids; firewalling of AC grids against cascading blackouts; compensation of AC system disturbances thanks to fast and flexible control of power flows. However, because power electronics devices typically have much smaller time constants than synchronous generators, it can bring the notorious "stiffness" issue in solving the system differential-algebraic equations during (conventional) transient simulation, causing super slow running or simulation diverging.

Therefore, a fast and robust (i.e., less sensitive to the stiffness issues) transient simulation tool is beneficial to power system operators for either post-fault study or preventive control purposes. However, mainstream off-the-shelf transient simulation software (proprietary products or open-source tools) can only meet the above need to some extent. Based on our tests, about seconds-level wall-clock time is typically required for a "10-sec" simulation of a mid-size power grid (e.g., a power grid with more than 10 high-order generator models plus exciters and governors), even on workstation-level computers.

EMTP, ATP-DRAW, and PSCAD/EMTDC are famous tools for (electromagnetic) transient simulation of HVDC systems [2]. However, due to the design of their internal numerical algorithms, their simulation speed for multi-machine power systems can be extremely slow or easily diverge when it considers high-order generator models. Other advanced solutions, such as products based on FPGA [3] and OpenMP [4], have been proposed by commercial companies. Unfortunately, the learning curves in hardware configuration or algorithm adaptation of those commercial products can be hard for junior researchers and students; the purchasing price can also be prohibitively high for university users.

In light of the above issues, other simulation approaches can be leveraged, e.g., modeling the power system components by electronic circuits. Previously, researchers have already employed circuit theory and tools in power system steady-state studies. For example, an electronic circuit theory-based robust solver is invented to solve three-phase power flow problems [5]. In [6], the *adjoint power flow* method is invented for the power flow feasibility study based on the electronic circuit theory. To our limited knowledge, so far, there is no public research reported for HVDC modeling in transient simulation using electronic circuit theory or tools.

This paper uses the electronic circuit simulator Xyce [7] for HVDC system modeling in power system transient simulation. Section II illustrates the theoretical foundations of electronic circuit simulation, including equation formulation and numerical solvers. Sections III and IV illustrate the electronic circuit modeling for three power system components: slack bus, constant power load, and a generic HVDC system. Simulation results are presented in Section V on a four-bus test system. Conclusions are given in the final section, with the next step indicated.

## II. FOUNDATIONS OF ELECTRONIC CIRCUIT SIMULATION

In this section, some foundations of circuit simulation theory are explained here based on [7].

*Modified nodal analysis* is usually applied to solve the circuit simulation problem, commonly used in the well-known SPICE program [7]. In the original KCL formulation, a circuit of $N$ nodes will have $N$-1 equations due to the KCL law, with $N$-1 voltage (state) variables (considering one node as a reference node with a known voltage). Meanwhile, currents between two nodes can be expressed as a function of the voltage drop, e.g., the current through a resistor satisfies $I=G*V$, where $G$ is the so-called "conductance". Moreover,

This study was supported by China State Grid Corporation technology project 5455HJ180018 in 2019.

the conductance may also depend on the voltage drop. Thus, the equation can be rewritten as $I=G(V)*V$. The above expression is called *conductance representation*. Electronic devices with conductance representations include resistors, capacitors, diodes, and transistors.

Unlike the original KCL formulation, the modified KCL formulation may incorporate non-Ohmic devices into the circuit equations via non-KCL equations and auxiliary variables (typically currents). To illustrate the above idea, two example circuits are described below: a linear circuit and a time-dependent circuit.

### A. Linear Circuits

By Ohm's law and KCL, the compact form equations for the circuit schematic in Fig. 1 are shown in Eq. (1).

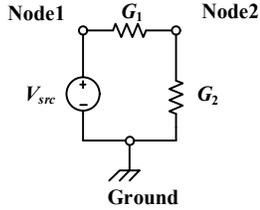

Figure 1. Example schematic for a linear circuit

$$\mathbf{f} := \begin{bmatrix} G_1 & -G_1 & 1 \\ -G_1 & G_1+G_2 & 0 \\ 1 & 0 & 0 \end{bmatrix} \begin{bmatrix} V_1 \\ V_2 \\ I_{Vsrc} \end{bmatrix} - \begin{bmatrix} 0 \\ 0 \\ V_{src} \end{bmatrix} = 0 \quad (1)$$

The Newton-type algorithms are typically utilized to solve the above circuit equations (denoting the state variable $\mathbf{x} = [V_1, V_2, I_{Vsrc}]$):

$$\mathbf{J}(x^k)\Delta\mathbf{x}^{k+1} = -\mathbf{f}(\mathbf{x}^k), \quad \mathbf{x}^{k+1} = \mathbf{x}^k + \Delta\mathbf{x}^{k+1} \quad (2)$$

In Eq. (2), the Jacobian matrix $\mathbf{J}$ is a 3-by-3 coefficient matrix and $\Delta\mathbf{x} = [\Delta V_1, \Delta V_2, \Delta I_{Vsrc}]$. The iteration repeats until the pre-defined stop criteria are reached, e.g., the uniform-norm $\|\Delta\mathbf{x}\|$ becomes less than 0.001.

### B. Time-Dependent Circuits

Electronic circuits can contain time-dependent elements, and many of those elements (e.g., capacitors and inductors) are described by ordinary differential equations (ODEs) with time-derivative terms. For example, the current through a constant capacitor is given by:

$$I_C = \frac{dq}{dt}, \quad q = C \cdot V \quad (3)$$

Then, the circuit equations will be a set of differential-algebraic equations (DAEs). Most time-dependent circuit equations can be transformed to the so-called *index one* DAEs, and the techniques for solving this kind of DAEs have been well studied. The DAEs can be described by Eq. (4):

$$\mathbf{f}\left(\mathbf{x}, \frac{d\mathbf{x}}{dt}, t\right) = \mathbf{0} \quad (4)$$

For example, the derived DAEs in compact form for the circuit schematic in Fig. 2 are shown in Eq. (5):

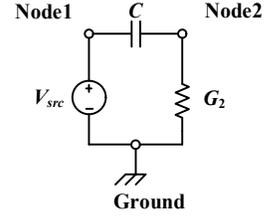

Figure 2. Example schematic for a time-dependent circuit.

$$\begin{bmatrix} 0 & 0 & 1 \\ 0 & G_2 & 0 \\ 1 & 0 & 0 \end{bmatrix} \begin{bmatrix} V_1^{i+1} \\ V_2^{i+1} \\ I_{Vsrc}^{i+1} \end{bmatrix} + \begin{bmatrix} C & -C & 1 \\ -C & C & 0 \\ 0 & 0 & 0 \end{bmatrix} \frac{d}{dt}\begin{bmatrix} V_1^{i+1} \\ V_2^{i+1} \\ I_{Vsrc}^{i+1} \end{bmatrix} + \begin{bmatrix} 0 \\ 0 \\ -V_{src} \end{bmatrix} = 0 \quad (5)$$

where the superscript $i$ stands for the $i$-th iteration. Using the backward differentiation formula (BDF) in (6), the DAEs can be rewritten as Eq. (7), where $h$ is the integration time step.

$$\frac{dV_{12}}{dt} = (V_{12}^{i+1} - V_{12}^i)/h \quad (6)$$

$$\frac{1}{h}\begin{bmatrix} C & -C & 1+h \\ -C & G_2h+C & 0 \\ h & 0 & 0 \end{bmatrix} \begin{bmatrix} V_1^{i+1} \\ V_2^{i+1} \\ I_{Vsrc}^{i+1} \end{bmatrix} + \frac{1}{h}\begin{bmatrix} C & -C & 1 \\ -C & C & 0 \\ 0 & 0 & 0 \end{bmatrix} \begin{bmatrix} V_1^i \\ V_2^i \\ I_{Vsrc}^i \end{bmatrix} = \begin{bmatrix} 0 \\ 0 \\ V_{src} \end{bmatrix} \quad (7)$$

Special techniques have been invented for accelerating the electronic circuit simulation, e.g., inexact Jacobian computation, state condensation, and variable-step integration. More details can be found in [7].

Besides, like other circuit simulators, Xyce uses the SPICE language to describe the *circuit netlist*. For example, the following sentence defines a constant voltage source of 1.20 V in-between "Node1" and "Node2" with the name "Vsrc" (in Xyce, the first letter "V" and "B" stands for voltage source and current source respectively):

**Vsrc Node1 Node2 1.20V**

For other grammar and usages of the SPICE language in Xyce, readers can refer to [8].

### III. MODELING OF POWER SYSTEM COMPONENTS IN XYCE

By default, the electronic circuit simulator does not accept complex numbers. However, real and imaginary parts of a power system quantity can be split into two circuit groups (actually, the two groups are still electronically connected). Xyce provides three different modeling formats to describe power systems components: PQP (means PQ Polar), PQR (PQ Rectangular), and I=YV (real and imaginary parts of voltage and current) [9]. The I=YV format uses the rectangular form of a complex number, whose solution variables are the real and imaginary parts of the voltage (e.g., "VR" and "VI") and current (e.g., "IR" and "II").

One basic idea of modeling power system components by the circuit simulator is: each "power system quantity" will be mimicked by an "electronic voltage or current signal" (in most cases, voltage signal), which is illustrated in the following paragraphs. Besides, similar to the *object* or *subroutine* concept in certain advanced programming languages, Xyce uses "subcircuit" (**SUBCKT**) to encapsulate a user-defined model (device).

### A. Slack Bus Model

The slack bus (steady-state generator) circuit schematic is shown in Fig. 3(a) (only shows the real part). In this subcircuit, two ideal voltage sources are connected in series. The top one is connected to the *actual bus* node denoted as 'Bus1R' with |V|=1.0507V. Due to the naming convention of Xyce, the nodes are named in a pattern of 'BusXXR' (real part signal) and 'BusXXI' (imaginary part signal), where 'XX' can be either a number or string. Note that, by the Xyce rule, a zero-value ideal voltage source represents the ideal ampere meter. Its positive end is placed nearby the ground to satisfy the "generator-convention" in power systems, i.e., *positive* when *injecting* power into the *bus*. Its SPICE-style netlist code is given in Appendix. A.

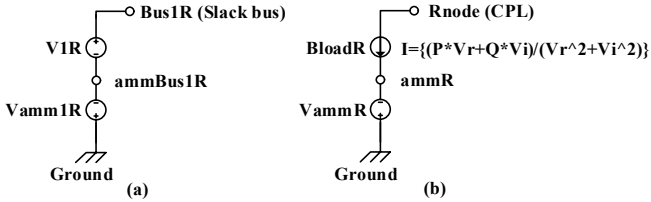

Figure 3. Example schematic (only real parts) for (a) slack bus (b) constant power load

### B. Constant Power Load (CPL) Model

The circuit schematic is shown in Fig. 3(b) (only shows the real part). The ampere meter is defined so that positive power flows into the load. To express the complex current by the associated real and reactive power injections (per unit) in the I=YV format, Eq. (8) is utilized:

$$I = I_r + jI_i = \left(\frac{S}{V}\right)^* = \frac{(PV_r + QV_i)}{V_r^2 + V_i^2} + j\frac{(PV_i - QV_r)}{V_r^2 + V_i^2} \quad (8)$$

The main part of the subcircuit netlist is presented in Appendix. B. Note that a "limiter" is added in the subcircuit code to prevent drastic current change during initializations. For example, if a higher-level netlist program needs to define a constant power load, the encapsulated CPL model can be defined as follows:

**Xload1 bus1R bus1I CPL PARAMS: P=0.9 Q=0.49**

## IV. MODELING OF HVDC IN XYCE

There are different HVDC models based on the different physical configurations of power electronic devices, filters, and controller designs. This section considers the LCC (line-commutated current-sourced converters) HVDC (VSC-type HVDC can also be modeled similarly by the methodology introduced in this paper). The typical one-line diagram of a monopolar back-to-back HVDC system is shown in Fig. 4.

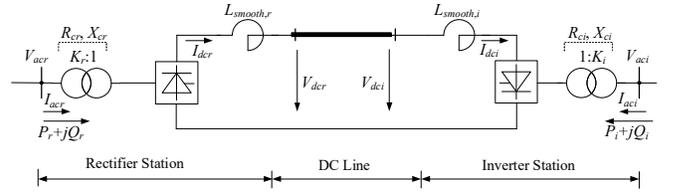

Figure 4. The one-line diagram of a monopolar HVDC system

All HVDC systems, whether two-terminal, multi-terminal, or back-to-back, have similarly designed AC/DC converter stations. In the system stability study involving HVDC converters, the interactions of the converter with the AC system are extremely important and complex: if the AC voltage at a converter bus is perturbed, there will be an immediate change in the DC voltage on the converter. This paper models the converter by the "average value model" for the power system (electromechanical) transient stability study.

Due to this paper's scope, the following points are not considered: 1) does not intend to handle unbalance system and unbalanced faults events; 2) does not consider secondary systems, such as protection relays, lightning arresters, etc.

Fig. 4 depicts a functional one-line diagram of a monopolar, two-terminal HVDC system. The subscript "*r*" represents the physics quantities of the rectifier side, and "*i*" represents the physics quantities of the inverter side.

### A. HVDC Line Model

In the transient stability study, each HVDC transmission line can be modeled as an equivalent resistor and an equivalent inductor, as shown in Fig. 5.

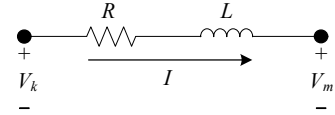

Figure 5. HVDC Transmission Line Model

The HVDC transmission line equation is shown in Eq. (9). The unknown variable here is the DC current *I*.

$$V_k - V_m - RI = L\frac{dI}{dt} \quad (9)$$

### B. Equivalent Circuit Model

In the transient stability study, by defining the following quantities in Eq. (10) to (12), the HVDC primary system can be conceptually represented by two (nonlinearly coupled) Thevenin-equivalent circuits, as shown in Fig. 6, where the rectifier, the HVDC line, and the inverter are modeled.

$$L_r = \frac{\text{DC Line Indcutance}}{2} + L_{smooth,r} + \frac{1.75 N_{rec} X_{cr}}{\omega} \quad (10)$$

$$L_i = \frac{\text{DC Line Indcutance}}{2} + L_{smooth,i} + \frac{1.75 N_{inv} X_{ci}}{\omega} \quad (11)$$

$$R_r = R_i = \text{DC Line Resistance}/2, \ X_{dc} = \omega(L_r + L_i) \quad (12)$$

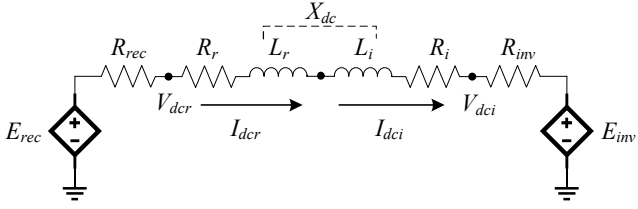

Figure 6.  HVDC (Thevenin) equivalent circuit model

The rectifier side algebraic equations are:

$$V_{dcr} = E_{rec} - N_{rec}(R_{rec}I_{dcr}), E_{rec} = \frac{3\sqrt{2}}{\pi}\frac{V_{acr}}{K_r}\cos\alpha$$
$$R_{rec} = \frac{3X_{cr}}{\pi} + 2R_{cr}, \mu = \cos^{-1}(\cos\alpha - \frac{\sqrt{2}I_{dcr}X_{cr}}{E_{acr}}) - \alpha \quad (13)$$

where $\alpha$, $\mu$, $N_{rec}$ are respectively the firing angle, the commutation overlap angle, and the number of commutation bridges at the rectifier side.

Similarly, the inverter side algebraic equations are:

$$V_{dci} = E_{inv} + N_{inv}(R_{inv}I_{dci}), E_{inv} = \frac{3\sqrt{2}}{\pi}\frac{V_{aci}}{K_i}\cos\beta$$
$$R_{inv} = \frac{3X_{ci}}{\pi} + 2R_{ci}, \beta = \cos^{-1}(\cos\gamma - \frac{\sqrt{2}I_{dci}X_{ci}}{E_{aci}}) \quad (14)$$

where $\gamma$, $\mu$, $N_{inv}$ are respectively the arc-extinction angle, the commutation overlap angle, and the number of commutation bridges at the inverter side, and $\beta := \gamma + \mu$ by its definition.

Finally, the modeling of HVDC in Xyce is illustrated in Fig. 7. Note that the detailed HVDC modeling in this paper is based on the "CHVDC2" structure in [10], of which a generic HVDC transfer function block diagram is shown in Fig. 8.

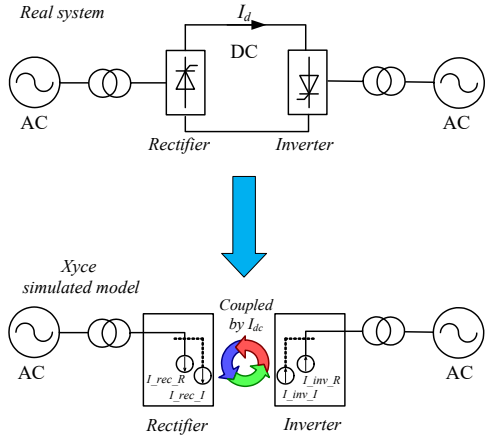

Figure 7.  The conceptual illustration of electronic-circuit modeling of HVDC (two-terminal, monopolar) in Xyce

### C. HVDC Control System Model

There are no standardized control models applicable to all schemes of HVDC. However, HVDC systems from various manufacturers share a common element, viz. control of the current, power, and firing (or extinction) angle of each converter. Also, the controller can be categorized as the inner-loop controller and the outer-loop controller. For example, one kind of inner-loop control can be found in Fig. 8, where *Idc_ref* (the DC current command) is impacted by AC side voltage *Vac* via complex logic.

Regarding outer-loop control, there can be various designs depending on different goals or modes of system operation. For example, the *Power and Current* control mode takes the active power or current as the control parameter. The power command is divided by the filtered DC voltage to yield the DC current command *Idc_ref*. Besides, the *Extinction Angle control* can be (simultaneously) used. Since the paper focuses on the modeling and simulation aspect of HVDC per se, details for the outer-loop controller are omitted here. More details can be found in [10].

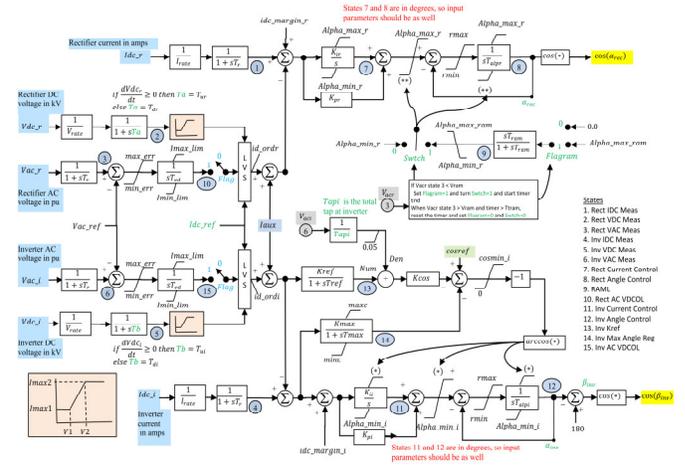

Figure 8.  A generic transfer function block diagram for HVDC [10]

### V. CASE STUDY

In this section, a four-bus, two-area power system is built in the Xyce environment with an HVDC link between bus-1 and bus-2 (as shown in Fig. 9). Bus-2 is connected with an "infinity bus" (i.e., $V=1.0\angle 0°$pu), which represents the external grid. The high-order model ("GENROU") with exciter and governor is adopted for the synchronous generator.

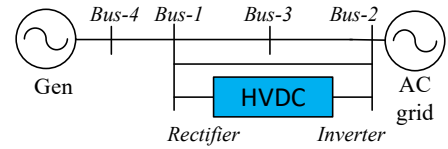

Figure 9.  A four-bus, two-area power system with HVDC

### A. Simulation Settings and Parameters

All the generators use high-order models considering exciters, governors, and saturation effects. Simulation settings and partial system parameters are listed in Tables I to III.

TABLE I.  SYSTEM PARAMETERS AND SIMULATION SETTINGS

| Xyce integral time step | 10ms | $x_{12}$ | 0.2pu |
|---|---|---|---|
| Simulation length | 200s | $x_{13}$ | 0.1pu |
| Xyce RelTolerance | 1e-1 | $x_{14}$ | 0.1pu |
| Xyce AbsTolerance | 1e-3 | $x_{23}$ | 0.1pu |
| $V_2$ magnitude | 1.0pu | $V_4$ magnitude | 1.0971pu |

TABLE II. HVDC PARAMETERS

| Tap | 1.0 | $X_{cr}$, $X_{ci}$ | 0.2 |
|---|---|---|---|
| $Idc\_ref$ | 1.0pu | $N_{rec}$, $N_{inv}$ | 6 |
| $Vac\_ref$ | 1.0pu | $P_{dc\_rec}$ | 0.8 |
| $X_{dc}$ | 0.111pu | $P_{dc\_inv}$ | 0.7 |

TABLE III. GENERATOR PARAMETERS

| $T_d'$ | 7s | $x_l$ | 0.15pu |
|---|---|---|---|
| $T_d''$ | 0.03s | $x_d$ | 2.1pu |
| $T_q'$ | 0.75s | $x_d'$ | 0.2pu |
| $T_q''$ | 0.05s | $x_d''$ | 0.18pu |
| $x_q$ | 0.5pu | $x_q'$ | 0.25pu |
| $r_a$ | 0.0pu | $x_q''$ | 0.18pu |
| $H$ | 3s | $D$ | 0.5pu |

### B. Simulation Results

#### 1) Without Extinction Angle Control

Let the inverter side not consider the extinction angle .firing/extinction angles are depicted in Fig. 10 and 11. The wall clock time cost is about 4.4 seconds for a "200sec" simulation length.

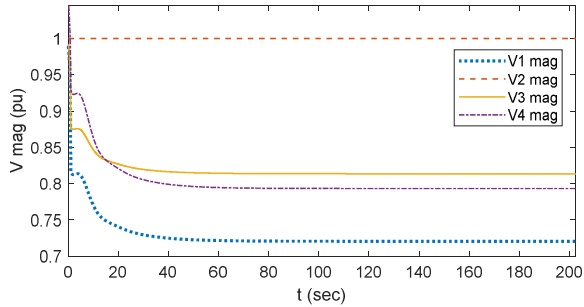

Figure 10. The bus voltages without extinction angle control

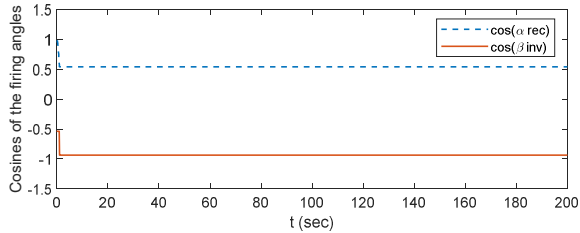

Figure 11. The firing/extinction angles without extinction angle control

It can be observed that when there is no extinction angle control, the voltage magnitudes at the rectifier side are relatively low due to the large reactive power supplied from the infinity system side.

#### 2) With Extinction Angle Control

Let the inverter side consider the extinction angle control and set $\beta_{ref} = 20°$. The responses of the bus voltages and firing/extinction angles are depicted in Fig. 12 and 13. The wall clock time cost is about 4.9 seconds for a "200sec" simulation length.

It can be observed that when the extinction angle control is adopted at the inverter side, the voltage magnitudes of buses at the rectifier side can finally increase to 1.0pu or above.

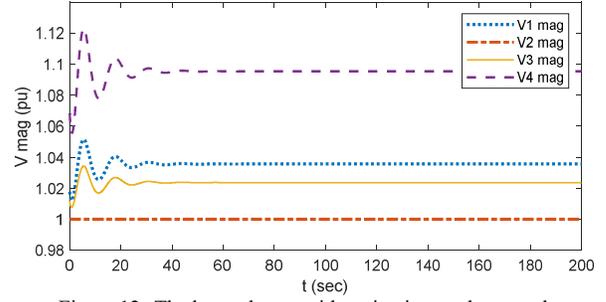

Figure 12. The bus voltages with extinction angle control

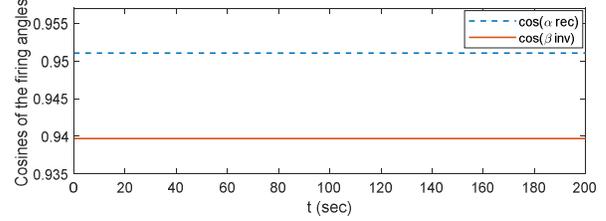

Figure 13. The firing/extinction angles with extinction angle control

## VI. CONCLUSION

In this paper, a circuit simulator-based HVDC modeling method for power system transient simulation is proposed. The case study results for a four-bus system demonstrate the method's applicability and computational speed. The next step is implementing other Xyce models for practical HVDC applications, e.g., the protective relays, filter banks, and multi-terminal HVDCs.

## APPENDIX: CODE SNIPPETS OF XYCE

### A. Slack Bus Modeling

```
**********************************************
* Bus 1 is the slack bus, VMag=1.0507, Vtheta=0.51deg*
**********************************************
V1R  Bus1R ammBus1R 1.0507V
V1I  Bus1I  ammBus1I 0.00935V
VammR 0 ammBus1R 0V
VammI 0 ammBus1I  0V
```

### B. CPL Modeling

```
**********************************************
* Constant Power Load                         *
**********************************************
.SUBCKT CPL RNode INode PARAMS: P=0.5 Q=0.0 CurrLim=1000
……
BloadR RNode ammR
+
I={limit((P*V(RNode)+Q*V(INode))/(V(RNode)*V(RNode)
+V(INode)*V(INode)), -CurrLim,CurrLim)}
……
BloadI INode ammI
+
I={limit((P*V(INode)-Q*V(RNode))/(V(RNode)*V(RNode)+V(INode)*V(INode)), -CurrLim, CurrLim)}
.ENDS
```

### C. HVDC Modeling

```
*****************************************
* Subcircuit definition for the HVDC CHVDC2 diagram
*****************************************
* Output cos_alpha_rec
* Output cos_beta_inv
.SUBCKT CHVDC2_Module Vac_Rec_r Vac_Inv_r
Vac_Rec_i Vac_Inv_i
+ PARAMS: I_rate=1 Tr=0.005 V_rate=1 Tur=1 Tdr=1
V1=1 V2=1 Imax1=1 Imax2=1
+ Alpha_min_r=0.21 Alpha_max_r=0.35 rmin=0 rmax=1
Talpr=1 Alpha_max_ram=0.21, Tap=1.0 Xc=0.2 Gdc =0.111
Nbr=6
……
BVac_rec  Vac_r 0
V={SQRT(V(Vac_Rec_r)*V(Vac_Rec_r)+V(Vac_Rec_i)*V(Vac_Rec_i))}
BVac_inv  Vac_i 0
V={SQRT(V(Vac_Inv_r)*V(Vac_Inv_r)+V(Vac_Inv_i)*V(Vac_Inv_i))}
……
BVdc_rec  Vdc_rec 0
V={Nbr*(3/PI)*(SQRT(2)*V(Eeq_rec)*V(cos_alpha_rec)-Xc*V(Idc))}
BVdc_inv  Vdc_inv 0
V={Nbr*(3/PI)*(SQRT(2)*V(Eeq_rec)*V(cos_beta_inv)-Xc*V(Idc))}
……
* (Rectifier)
VammR_rec ammR_rec 0 0V
VammI_rec ammI_rec 0 0V
BloadR_rec Vac_Rec_r ammR_rec
+ I={limit( (Pdc_rec*V(Vac_Rec_r)+
V(Qdc_rec)*V(Vac_Rec_i))/(V(Vac_Rec_r)*V(Vac_Rec_r)
+V(Vac_Rec_i)*V(Vac_Rec_i)), -1000,1000)}
BloadI_rec Vac_Rec_i ammI_rec
+ I={limit( (Pdc_rec*V(Vac_Rec_i)-
V(Qdc_rec)*V(Vac_Rec_r))/(V(Vac_Rec_r)*V(Vac_Rec_r)
+V(Vac_Rec_i)*V(Vac_Rec_i)), -1000,1000)}
* (Inverter)
VammR_inv ammR_inv 0 0V
VammI_inv ammI_inv 0 0V
BloadR_inv Vac_Inv_r ammR_inv
+ I={limit( (-Pdc_inv*V(Vac_Inv_r)-
V(Qdc_inv)*V(Vac_Inv_i))/(V(Vac_Inv_r)*V(Vac_Inv_r)+
V(Vac_Inv_i)*V(Vac_Inv_i)), -1000,1000)}
BloadI_inv Vac_Inv_i ammI_inv
+ I={limit( (-Pdc_inv*V(Vac_Inv_i)+
V(Qdc_inv)*V(Vac_Inv_r))/(V(Vac_Inv_r)*V(Vac_Inv_r)+
V(Vac_Inv_i)*V(Vac_Inv_i)), -1000,1000)}
.ENDS
```